# Coexistence of multiuser entanglement distribution and classical light in optical fiber network with a semiconductor chip

Xu Jing,[1,2,†] Cheng Qian,[1,†] Hu Nian,[1] Chenquan Wang,[2] Jie Tang,[2] Xiaowen Gu,[2] Yuechan Kong,[2] Tangsheng Chen,[2] Yichen Liu,[3,4,7] Chong Sheng,[3,8] Dong Jiang,[5,9] Bin Niu,[2,10] and Liangliang Lu[1,3,6,*]

[1]*Key Laboratory of Optoelectronic Technology of Jiangsu Province, School of Physical Science and Technology, Nanjing Normal University, Nanjing 210023, China*
[2]*National Key Laboratory of Solid-State Microwave Devices and Circuits, Nanjing Chip Valley Industrial Technology Institute, Nanjing Electronic Devices Institute, Nanjing, 210016, China*
[3]*National Laboratory of Solid-State Microstructures, Nanjing University, Nanjing 210093, China*
[4]*Research Center for Quantum Optics and Quantum Communication, School of Science, Qingdao University of Technology, Qingdao 266520, China*
[5]*School of Internet, Anhui University, Hefei 230039, China*
[6]*Hefei National laboratory, Hefei 230088, China*
[†]*These authors contributed equally to this work*
[7]*e-mail: liuyichen@qut.edu.cn*
[8]*e-mail: csheng@nju.edu.cn*
[9]*e-mail: jiangd@nju.edu.cn*
[10]*e-mail: niubin_1@126.com*
[*]*e-mail: lianglianglu@nju.edu.cn*

**Abstract:** Building communication links among multiple users in a scalable and robust way is a key objective in achieving large-scale quantum networks. In realistic scenario, noise from the coexisting classical light is inevitable and can ultimately disrupt the entanglement. The previous significant fully connected multiuser entanglement distribution experiments are conducted using dark fiber links and there is no explicit relation between the entanglement degradations induced by classical noise and its error rate. Here we fabricate a semiconductor chip with a high figure-of-merit modal overlap to directly generate broadband polarization entanglement. Our monolithic source maintains polarization entanglement fidelity above 96% for 42 nm bandwidth with a brightness of $1.2*10^7$ Hz/mW. We perform a continuously working quantum entanglement distribution among three users coexisting with classical light. Under finite-key analysis, we establish secure keys and enable images encryption as well as quantum secret sharing between users. Our work paves the way for practical multiparty quantum communication with integrated photonic architecture compatible with real-world fiber optical communication network.

## 1. Introduction

The development of quantum network (QN) is crucial for exploring applications beyond its classical counterpart, such as distributed and blind computing [1, 2], enhanced sensing [3, 4], and ultra-secure communications [5, 6]. To date, QNs of various topologies have been developed with trusted relays or active/passive routing [7-17]. Among them, fully and simultaneously connected QN architecture is desirable due to its practicality and scalability [12-16]. In this architecture, because of the nonlocal properties of quantum states distributed in networks, dedicated entanglement between any pair of users can be established even without a direct optical link [17]. Currently, bring QNs closer to practice is essential for scaling the connection of quantum nodes. A promising approach for deploying scalable QNs is to leverage off-the-shelf fiber infrastructures and technologies, and transmit both quantum and classical signals in shared fibers, as this would efficiently reduce the costs.

Up to now, a series of quantum-classical coexistence investigations have been implemented in fiber-based quantum communications [18-26]. However, most of these studies have focused solely on the point-to-point quantum key distribution (QKD) protocol, and the coexistence of multiuser entanglement distribution and classical light has not yet been fully exploited. The main challenge for hybrid quantum-classical communications is the strong noise induced by high power classical light. Similar to classical communication, interaction with noise will eventually reduce the channel capacity, becoming a limiting factor in terms of both rate and distance in quantum communications. Therefore, it is crucial to design advanced protocols, which can withstand large amounts of noise [27].

High dimensional (HD) entanglement is often considered to have the potential to provide stronger violations of Bell inequalities and better noise tolerance in quantum communications [28, 29]. Despite all this, the high noise resistance of HD entanglement has been recently demonstrated in realistic scenarios [30]. This is primarily because the HD quantum states are difficult to control and measure, and the increase of system dimensions often introduces additional noise [31, 32]. To implement entanglement-based HDQKD more effectively, a simultaneous multiple subspace coding protocol is proposed, enabling a secret key even in extremely noisy conditions [33]. Then the theoretical prediction has been verified in a proof-of-principle implementation using path entanglement with multi-outcome measurements [34]. Nonetheless, the scalability of this approach is limited due to the increase of detector number and the difficulty in achieving phase stability. Fortunately, the subspace noise-resistant protocol has been realized in [35] using polarization-frequency hyperentanglement is practically deployable and naturally suitable for fiber QNs. This pioneer scheme offers three advantages.

Firstly, the d×d-dimensional hyperentangled system can be divided into $l$ subspaces of size 2×2 (polarization entanglement), where d= $l$×2. The polarization qubit subspace can be used for coding and achieves the highest noise resistance [33], while the frequency qudit subspace is employed to dilute dimension-dependent noise. Secondly, polarization is easy to control and analyze without nested interferometers, and has mature automatic polarization feedback system. Additionally, polarization encoding has been operated steadily in installed fiber networks for several hours without active compensation [36]. Thirdly, the multiplexing of frequency degree of freedom, compatible with off-the-shelf fiber network infrastructures, can not only improve data rates in classical communications [37] but also provide wide connectivity for QNs.

Here, we leverage and expand upon these advantages to establish a fully connected three-user QN that operates in tandem with classical light, utilizing a high-efficiency semiconductor chip. First of all, the entangled state is generated via modal phase matching (PM) within the Bragg reflection waveguide (BRW) structure, employing AlGaAs/GaAs-based material. The quantum source achieves PM by utilizing bounded total internal reflection (TIR) modes formed between high- and low-index claddings, along with quasi-bounded BRW modes guided though transverse Bragg reflections at the interface between core and period claddings[38, 39]. Meanwhile, the source is chip-based, providing stable and alignment-free operations[40]. It is CMOS compatible and well-developed for various devices such as lasers and modulators [39, 41-46]. We show that, our monolithic source maintains polarization entanglement fidelity exceeding 96% across a 42 nm bandwidth, with a brightness of $1.2*10^7$ Hz/mW. Furthermore, the polarization entangled photons are partitioned and employed for the characterization of its increased resilience to noise. In our experiments, the noise is introduced by coexisting a classical light with quantum signals in the same fiber, which is extremely practical for installed fiber networks. Subsequently, we achieve a fully connected three-user quantum communication in a noisy network by employing three pairs of polarization entangled photons. Each user analyzes and decodes the polarization qubits using two unbalanced polarization-maintaining interferometers (UPMIs) and a single detector via time multiplexing. Finally, the generated secure keys are used for images encryption and quantum secret sharing (QSS). Overall, our results, combined with the utilization of subspace coding technology, present a promising pathway for advanced QN applications in deployed fibers.

**2. Noise models and experimental verifications**

First, let us briefly review the protocol developed in [35]. Assuming entanglement in both polarization and frequency degrees, the ideal entangled state can be expressed as [47]:

$$\rho_1 = |\Psi^+\rangle\langle\Psi^+| \otimes |\psi\rangle\langle\psi| \qquad (1)$$

where $|\Psi^+\rangle = \frac{1}{\sqrt{2}}(|HV\rangle + |VH\rangle)$ and $|\psi\rangle = \sum_{n=1}^{N}\frac{1}{\sqrt{N}}|\omega_{s,n}\rangle|\omega_{i,N-n}\rangle$, with $\omega_{s,n} = \omega_0 + n\Delta\omega$ and $\omega_{i,N-n} = \omega_N - n\Delta\omega$ represent signal and idler photons chosen from the spectrum such that the sum of $\omega_0\ and\ \omega_N$ equals to the pump frequency. $N$ is the total number of frequency-bin

pairs selected. $|H\rangle/|V\rangle$ represents the horizontally or vertically polarized states. As shown in Fig. 1a, $\rho_1$ transmits over noisy channels can be considered as pairs of frequency correlated photons pass through independent links and measured separately. We discuss now some of the most usual noise for qubits: amplitude damping and white noise. The influence of white noise on the state can be described as

$$\rho_2 = (1-p)|\Psi^+\rangle\langle\Psi^+| \otimes |\psi\rangle\langle\psi| + \frac{p}{(2\times N)^2} I_2 \otimes I_N \quad (2)$$

where $p$ is the noise portion and $I_2(I_N)$ represents the completely mixed state in the polarization (frequency) subspace. After trancing over the frequency, the state can be written as an isotropic state

$$\rho_2' = \sum_{n=1}^{N}\langle nn|\rho_2|nn\rangle = (1-p)|\Psi^+\rangle\langle\Psi^+| + \frac{p}{4N} I_2 \quad (3)$$

As can be seen, the white noise leads to accidental coincidences in experiments and can be equivalently divided by the frequency subspace dimension $N$. This is the physical origin of noise resistance in this protocol. By choosing appropriate frequency channels, the probability of measuring false coincidences between noise-noise photons or noise-signal photons is reduced, while the probability of measuring a true coincidence between correlated photons is essentially unchanged. Therefore, the additional quantum correlation in frequency can be used to distill the signal photons from noise photons. Moreover, physical noise for quantum systems with a quantum state are generally characterized by the Kraus representation[48, 49]. The amplitude damping noisy channel represents the dissipative interaction between the qubit and its environment. The corresponding Kraus operators are

$$M_0 = \begin{pmatrix} 1 & 0 \\ 0 & \sqrt{1-\gamma} \end{pmatrix}, \quad M_1 = \begin{pmatrix} 0 & \sqrt{\gamma_i} \\ 0 & 0 \end{pmatrix} \quad (4)$$

where $\gamma_i$ (i=1,2) denotes the noise portion. Therefore, the noise on $\rho_2'$ can be modeled by

$$\rho_2'' = \sum_{i,j=0}^{1} M_i \otimes M_j \, \rho_2' (M_i \otimes M_j)^\dagger \quad (5)$$

Given the above density matrix of a composite quantum system, the entanglement negativity [50, 51] between two bipartitions of the system is given as

$$\mathcal{N} = \frac{1}{2N(-1+p)-2p}(p + N\gamma_1 - Np\gamma_1 + N\gamma_2$$
$$-Np\gamma_2 - 2N\gamma_1\gamma_2 - p\gamma_1\gamma_2 + 2Np\gamma_1\gamma_2 - \sqrt{\Delta}) \quad (6)$$

where

$$\Delta = 4N^2(-1+p)^2 + (N+p-Np)^2\gamma_1^2 - 4N^2(-1+p)^2\gamma_2 + (N+p-Np)^2\gamma_2^2 + 2\gamma_1(-2N^2(-1+p)^2 + (N^2(-1+p)^2 + 2N(-1+p)p - p^2)\gamma_2) \quad (7)$$

For simplicity, in Fig. 1b we model the effects of amplitude damping noise with coefficients $\gamma_1 = \gamma_2 = \gamma \in [0.1, 0.3]$, and white noise with portion $p$ on such state. It is clear that the effect of noise can be significantly reduced by increasing $N$, which indicates quantum correlation is an effective resource to resist noise. It is worth mentioning that the simple noise models we analyzed here only serves as an example for why the increased noise resistance should be expected. We do not assume any noise model when quantitatively analyzing the realistic quantum-classical signals coexistence in fiber network.

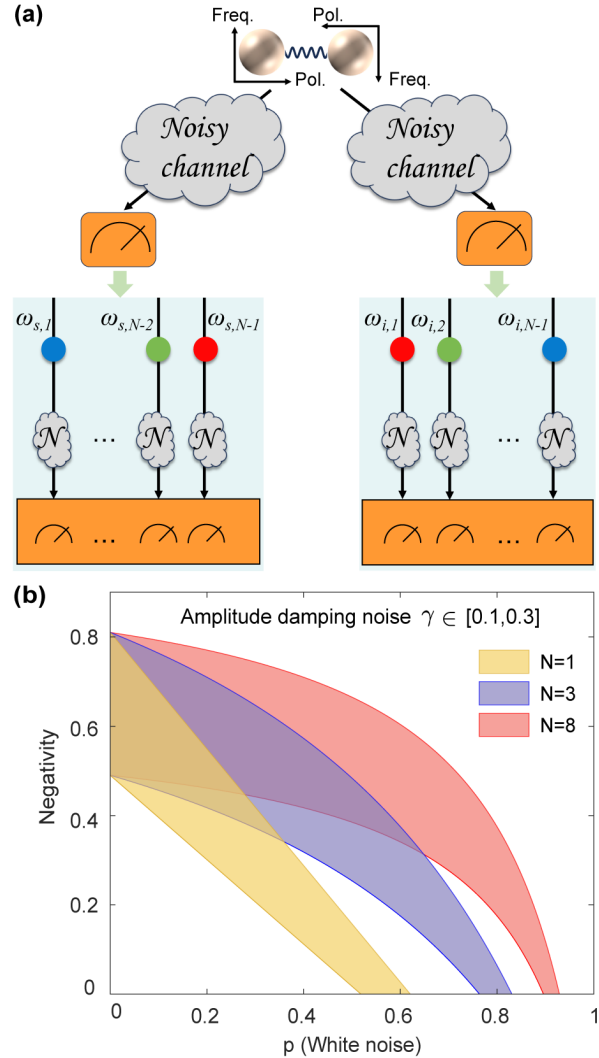

**Fig. 1.** Correlation-assisted quantum communication over noisy channels. (a) Using the frequency correlation, one of the polarization entangled photons with $N$ frequency bins passes through a noisy channel can be regarded as multiplexing into $N$ frequency-resolved channels and measured separately. (b) Simulations of entanglement negativity as a function of the white noise p with amplitude damping noise $\gamma$ in the range between 0.1-0.3 for three divisions. As the number of divisions increases, stronger entanglement can be observed, indicating a higher resilience to noise.

Fig. 2a shows the schematic of our experimental setup, containing three parts: photon-pair source, classical-quantum link and wavelength-selective switch (WSS), and polarization analysis module. In the source part, a single mode fiber-coupled continuous wave (CW) laser centered at 780.08 nm pumps the semiconductor BRW source. We split 1% power of the laser to the optical spectrum analyzer (OSA) for wavelength calibration and use a fiber polarization controller (PC) to adjust the pump polarization before injecting into the source. The source consists of a core layer sandwiched by six periods upper and lower Bragg layers and has a length of 4 mm. Broadband polarization entangled photon pairs can be generated directly through degenerated type-II spontaneous parametric down conversion (SPDC) due to the small material birefringence (see Supplementary Note 1 for detailed description). The lack of emitted time information of photons in the SPDC process leads to energy-time entanglement [52]. Light is coupled into and out from the chip by lens tapered fibers mounted on high-precision servo motors. The total insertion loss of the chip is about 11dB, including both input-output coupling loss and propagation loss in the chip for the TIR modes. To stabilize the pump transmission, we use the rejected pump light filtered by a 980/1550 dense wavelength division multiplexer (DWDM) as a feedback signal for the motor and adopt hill-climbing algorithm to optimize the coupling under unstable laboratory conditions. Due to energy conservation during the SPDC process, the polarization entangled photons exhibit anti-correlation in frequency. The correlated photons and an external noise source are then multiplexed into the same fiber link via DWDM. In the experiment, noise is introduced by co-transmitting a 1591.26 nm classical laser with quantum signals through a 3-km-long fiber, representing a realistic scenario for a quantum communication system operating on public DWDM networks. The coexistence of classical and quantum signals in the same fiber can lead to some problems that may affect the performance of quantum communication. The dominant sources of noise are crosstalk from the classical light when the used DWDM has insufficient isolation, and Raman scattering which occurs due to the inelastic photon-phonon interaction[53]. In our experiment, the wavelength of classical light is far away from that of quantum signals, therefore we can eliminate the non-adjacent channel crosstalk with the use of appropriate filters at the quantum receivers. However, the broadband Raman noise cannot be spectrally filtered out as it is in-band with the quantum signals. In order to obtain the Raman cross section for different wavelength, we refer to the measurement results obtained using a laser centred at 1555 nm in a standard single mode fiber (see Supplementary Note 3). By using a wavelength selective switch (WSS), we postselect six wavelength grids of the signal (range from 1554.1 nm to 1557.1 nm) and idler (range from 1563.4 nm to 1566.4 nm) spectra. Each selected grid has a full width half maximum (FWHM) bandwidth of 0.4 nm and wavelength interval of 0.6 nm. The polarization analysis module, along with each channel, comprises a quarter-wave plate, a half-wave plate, an UPMI and a single photon detector (SPD). In the interferometer, two orthogonal polarizations are connected to polarization-maintaining fibers of different lengths, converting them into arrival time information for the photons. We can consequently obtain the number of coincidence counts in 4 configurations in one data accumulation. The single-photon detection events are recorded using an FPGA-based timetag unit. From these timetag records, we extract both single counts and coincidence counts. We split the *d*-dimensional Hilbert space into *d/2* mutually exclusive polarization subspaces. In Fig. 2b, we show the 6×6 joint spectrum intensity for four different classical laser powers in the Z={H, V} and X={A, D} bases with an integration time of 120 s. We set the input pump power to 30 mW and a coincidence window of 0.5 ns. As expected, an increase in the power of the classical laser results in a higher number of noise photons, which randomly spread the whole selected spectra range, while the correlated signals distributed in the diagonal elements almost remain constant. From the project measurements on the two mutually unbiased bases, a lower bound of fidelity to the Bell state can be estimated[45]. Figure 2b gives the averaged lower bound fidelity of the six frequency subspaces shown diagonally. The decrease in the fidelity can provide an intuitive explanation for the state becoming more mixed in stronger noisy channels (see Supplementary Note 3 for detailed theoretical analysis). Fig. 2c shows the averaged quantum bit error rates (QBER) in Z and X bases of various division number for different classical power. The QBER is determined by using the raw coincidence counts in each basis: $QBER_z=(C_{HH}+C_{VV})/(C_{HH}+C_{HV}+C_{VH}+C_{VV})$ and $QBER_x=(C_{DA}+C_{AD})/(C_{DA}+C_{DD}+C_{AD}+C_{AA})$. To study the effects of coarse-grained frequency division, we sum up the 6×6 matrix to reconstruct 1×1, 2×2 and 3×3 matrices for the estimation of QBERs. The results show that as the number of mutually exclusive subspaces increases, QBERs decrease dramatically, especially over highly noisy quantum channels, indicating stronger noise resilience for larger *N* in QNs. Fig. 2d gives the lower bounded concurrence calculated by $C \geq V_{D/A} + V_{H/V} - 1$ [54] as a function of division number with four classical light powers. All the experimental results are in good agreement with theoretical predictions (colored solid lines) calculated through the analysis in Supplementary Note 3. The increased degree of entanglement for large *N* implies nonlocality distillation, which also has the potential application to improve the noise resistance in device-independent QKD [55].

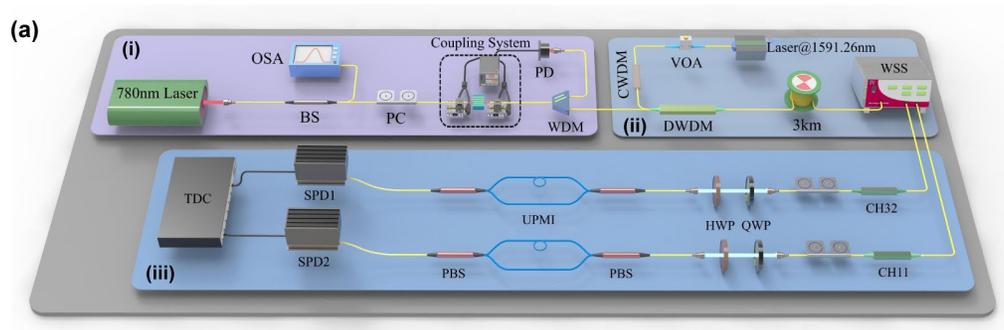
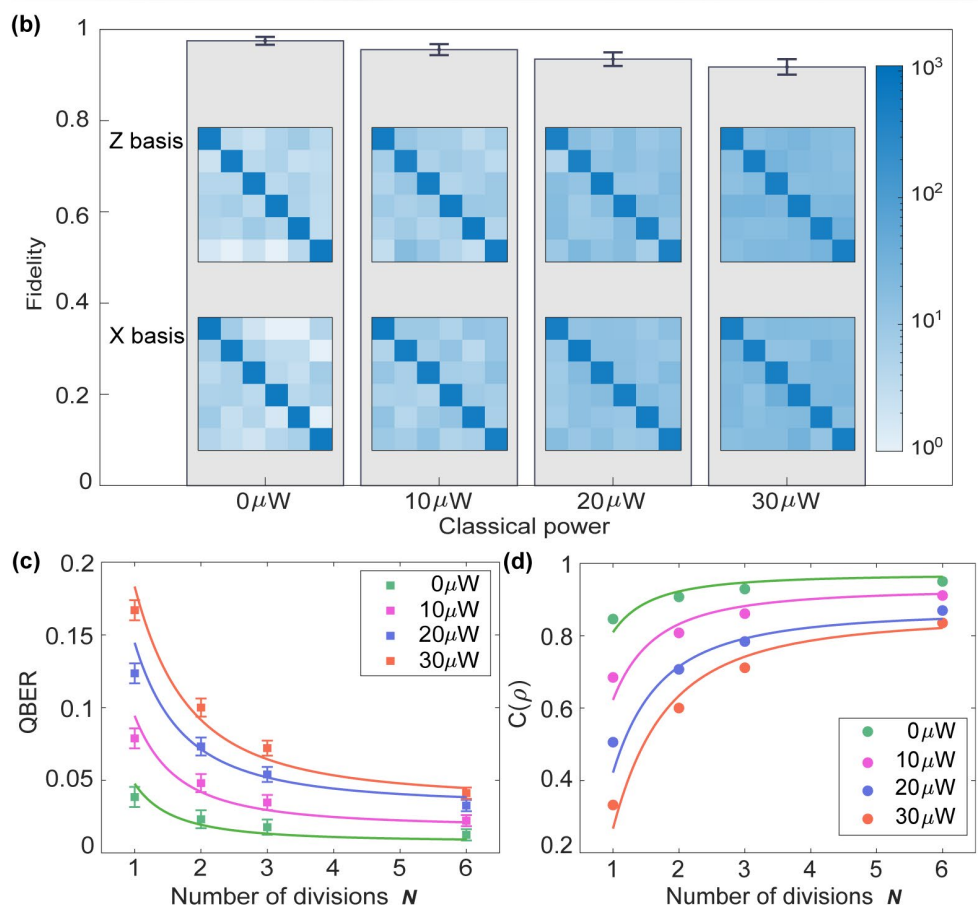

**Fig. 2.** Experimental setup and main results of noise resilience. (a) (i) Photon-pair generation. A continuous wave laser centered at a wavelength of 780.08nm is split by a 1:99 beam splitter (BS), wherein 1% of the light is sent to the optical spectrum analyzer (OSA) for wavelength detection. The remaining light is polarized and coupled into the semiconductor BRW by using a lensed fiber to generate broadband polarization entangled photon pairs. At the output, the pump laser is filtered by a 980/1550 wavelength division multiplexer and detected by a power detector as a feedback signal for the servo motor with hill-climbing algorithm to stabilize the fiber-chip coupling. (ii) Coexistence of quantum and classical light and wavelength demuxer. (iii) Analysis and detection. Each channel is equipped with a half-wave plate and a quarter-wave plate to project the photon onto HV or AD bases. The unbalanced polarization-maintaining interferometer (UPMI) introduces a polarization-dependent time delay to distinguish polarization states. Abbreviations of components: CW, continuous wave; PD, power detector; VOA, viable optical attenuator; CWDM, coarse wavelength division multiplexer; DWDM, dense wavelength division multiplexer; HWP, half-wave plate; QWP, quarter-wave plate; PBS, polarization beam splitter; SPD, single photon detector; TDC, time digital converter. (b) The lower bound to a Bell state $|\Psi^\pm\rangle$ are plotted for various powers of the classical signal at 1591.26 nm, along with normalized measurement matrices of six frequency subspaces. Each frequency-bin has a bandwidth of 0.4nm and spacing of 0.6 nm. The solid lines represent the theoretical curves. (c) and (d) respectively depict the averaged error rates and I-concurrence with different frequency-bin number against the classical signal power.

## 3. Main experimental results for multiuser quantum communication coexisting with classical light

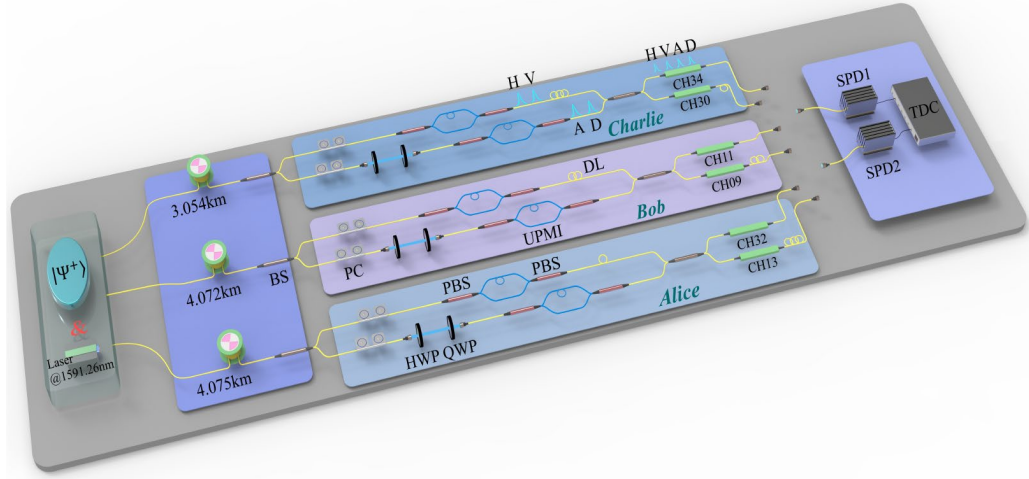

**Fig. 3.** Scheme of multiuser quantum entanglement distribution over noisy fiber networks. A broadband polarization entangled $|\Psi^+\rangle$ Bell state is multiplexed and distributed to three users. The spectrum is split into 6 channels to make each user receives two frequency channels, and therefore shares a polarization entangled pair with everyone else in the network. Then the classical laser located at 1591.26 nm is multiplexed with quantum signals into the same fiber to simulate the realistic scenario in installed fiber networks. At the receiver station, each user selects the basis passively and transfers polarization to time via two unbalanced polarization-maintaining interferometers (UPMIs). A time shift is introduced between vertically (diagonally) and horizontally (anti-diagonally) polarized photons. After that, a basis-dependent delay for Z and X is introduced to transfer the four polarization states to time-delayed states. The frequency-resolved detection can be realized by adding delays for each wavelength channel. The time/wavelength-division multiplexing technique adopted can cost-effectively reduce the detector number at per receiver to one.

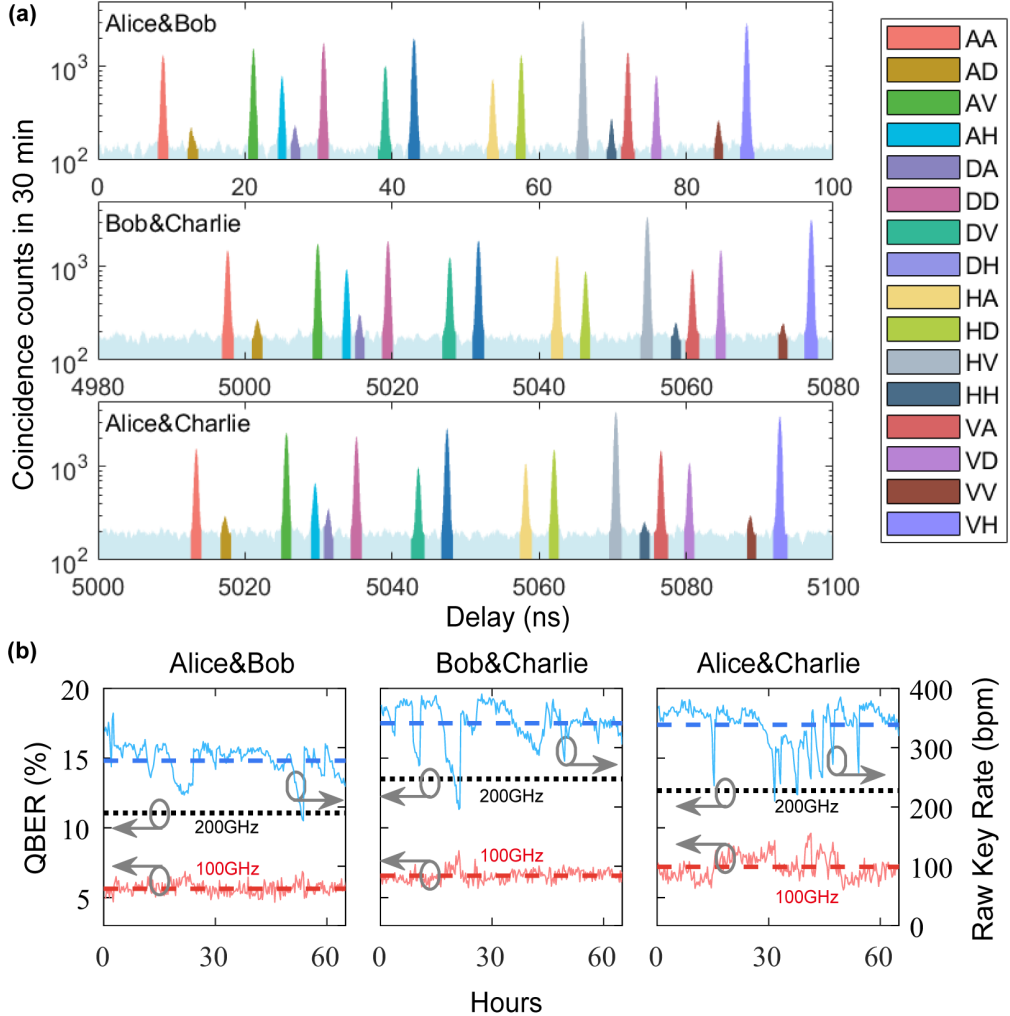

**Fig. 4.** Experimental results. (a) Photon time-correlation histograms correspond to three links by different delays for each real-time two-party connection. All the possible coincidences are indicated. (b) The quantum bit error rates (QBERs) and sifted key rates between Alice&Bob, Bob&Charlie, and Alice&Charlie in the noisy networks are measured for more than 65 hours. The averaged error rates for 100GHz DWDM are 5.63%, 6.58% and 7.21%, respectively. The QBERs of 200GHz DWDM are all above the 11% threshold and denoted as black dashed lines for comparison.

As a second series of measurements for a full use of the broad entanglement bandwidth and a complete demonstration of noise resilience in a quantum communication network, we perform entanglement distribution for a fully connected three-user network. As shown in Fig.3, the central server hosts a broadband polarization entangled photon pairs and select three channel pairs {CH32, CH13}, {CH11, CH09} and {CH34, CH30} via multiplexing and demultiplexing. Each user is assigned a combination of wavelengths as indicated in the detection module. We again use a classical laser to introduce noise to the fiber channel. The launch power of the laser in each fiber is set as -23dBm to reach the receiver sensitivity for 10Gbps classical signal. In the network, Alice is connected via a 4.075 km spool with a loss of 1.9 dB, Bob and Charlie use 4.072 km (1.8 dB) and 3.054 km (1.2 dB) spools, respectively. At the receiving stage, photons incident on the polarization analysis module (up: H/V, down: A/D) are passively split by a 50:50 BS for basis choice. For each basis, we use the UPMI to identify polarization by the relative arrival times of photons. An additional delay is added between the two bases to further distinguish the basis information (see Supplementary Note 4 for detailed description). We combine the two outputs by a 50:50 coupler to reduce the detector number at an expense of a 3 dB loss. The channel is distinguished by implementing wavelength-dependent time multiplexing as the method used in ref.[56]. Alternatively, one can combine the two single-mode fibers into one multi-mode output[57]. Due to the lack of an additional SPD, we run the two-user link separately, while the relative delay is added to make out different combinations of users. The time-correlation histograms corresponding to the highest recorded QBER for the three links are displayed in Fig. 4a. The simultaneous projection measurements on the Z and X bases are indicated by the marked sixteen coincidence combinations. Figure 4b shows the evolutions of the averaged QBERs and sifted key rates per minute with 100 GHz DWDMs in more than 65 hours from a laboratory test. Steep spikes in the figures are mainly caused by room-temperature variations, which result in disturbances to the coupling efficiency and the polarization states transmitted in the fibers. For comparison, we also recorded data for half an hour using 200 GHz DWDMs. The averaged QBERs are 11.07%, 13.51% and 12.67% respectively (denoted by black dashed lines), which are all above the 11% QBER threshold and much larger than those of 100 GHz DWDMs {5.63%, 6.58%, 7.21%}. These results clearly demonstrate the practical feasibility of that quantum correlation assisted multi-user entanglement distribution in real-world telecommunication fiber networks.

## 4. Secure key generation

To maximize the finite secret key length, we follow the security analysis and use the formula defined in ref.[58]. For standard QKD security definition, if a protocol is $\varepsilon_1$ correct and $\varepsilon_2$ secret, we can define the protocol is $\varepsilon_{QKD}$ secure with $0<\varepsilon_1+\varepsilon_2 \leq \varepsilon_{QKD}$, satisfying

$$2^{-t} + 2\varepsilon_{pe}(\nu,\xi) + \varepsilon_{pa}(\nu) \leq \varepsilon_{QKD} \tag{8}$$

We assume the finite key length of our implementation is $l = \lfloor \alpha m \rfloor$, where m is the sifted key length and $0<\alpha<1$. The error functions due to parameter estimation (pe) and privacy amplification (pa) can be defined as

$\varepsilon_{pe}(\nu,\xi) =$

$$\sqrt{\exp\left[-\frac{2m\beta\xi^2}{m(1-\beta)+1}\right] + \exp\left[-2\left(\frac{1}{m(\bar{E}+\xi)+1} + \frac{1}{m-m(\bar{E}+\xi)+1}\right)\right] \times [m^2(1-\beta)^2(\nu-\xi)^2 - 1]}$$

(9)

$$\varepsilon_{pa}(\nu) = \frac{1}{2}\sqrt{2^{-n[1-h_2(\overline{E})]+\overline{f}h_2(\overline{E})n+t+l}} \tag{10}$$

and

$$0 < \xi < \nu < \frac{1}{2} - \overline{E} \tag{11}$$

where the correctness error is $2^{-t} = 10^{-(s+2)}$, and security parameter is $\varepsilon_{QKD} = 10^{-s}$. f is the error correction efficiency shown in Supplementary Note 5. Therefore, we numerically maximize $l$ by optimizing $\{\alpha, \beta, \nu, \xi\}$ while conforming to the inequality formula (8) with fixed block length and security level s=9. We numerically find the parameter values {0.262, 0.080, 0.015, 0.013} for Alice & Bob, {0.149 0.100 0.014 0.012} for Bob & Charlie and {0.191,0.09,0.015,0.013} for Alice & Charlie, respectively. The allowed longest keys are 388.56 kb, 199.72 kb and 254.63 kb respectively.

## 5. Image encryption and three-user QSS

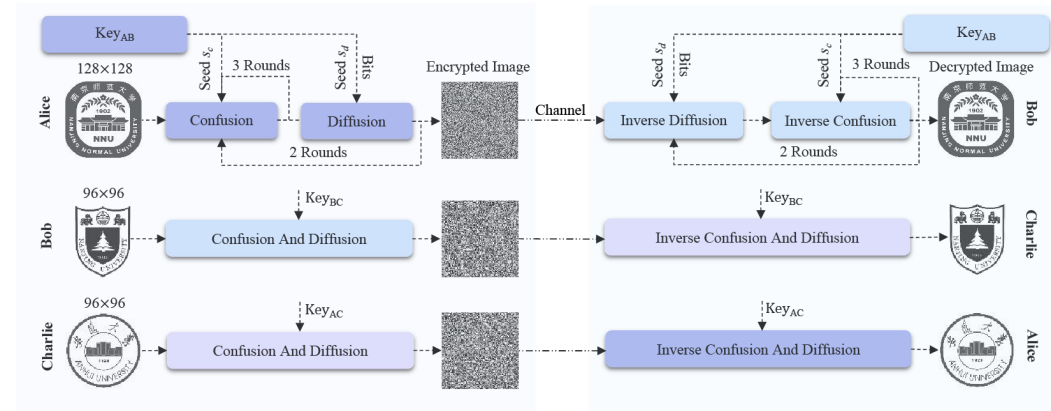

**Fig. 5.** Experimental procedure and results of multi-user image encryption and decryption. The sender performs three rounds of confusion operations, followed by two rounds of diffusion and confusion operations using the secret keys to encrypt the original image, and sends the encrypted image to the receiver via the classical channel. The receiver decrypts the obtained image by performing the inverse transformation of encryption.

With the use of the generated secure keys, we realize image encryption, transmission, and decryption between each pair of users. Diverging from the commonly employed approach of directly encrypting images using XOR operations, our experiment incorporates the Shannon confusion-diffusion architecture to safeguard the encrypted images against image-specific attacks, such as cropping attacks [59, 60]. In this architecture, confusion involves shuffling the pixel positions without altering their values, while diffusion entails sequentially modifying the pixel values using secret keys (for further details, please refer to Supplementary Note 6). In our experiment, as illustrated in Fig. 5, each sender performs three rounds of confusion operations, followed by two rounds of diffusion and confusion operations to encrypt the original image.

The encrypted image is then transmitted to the receiver via the classical channel. The receiver performs the inverse transformation of encryption to obtain the decrypted image.

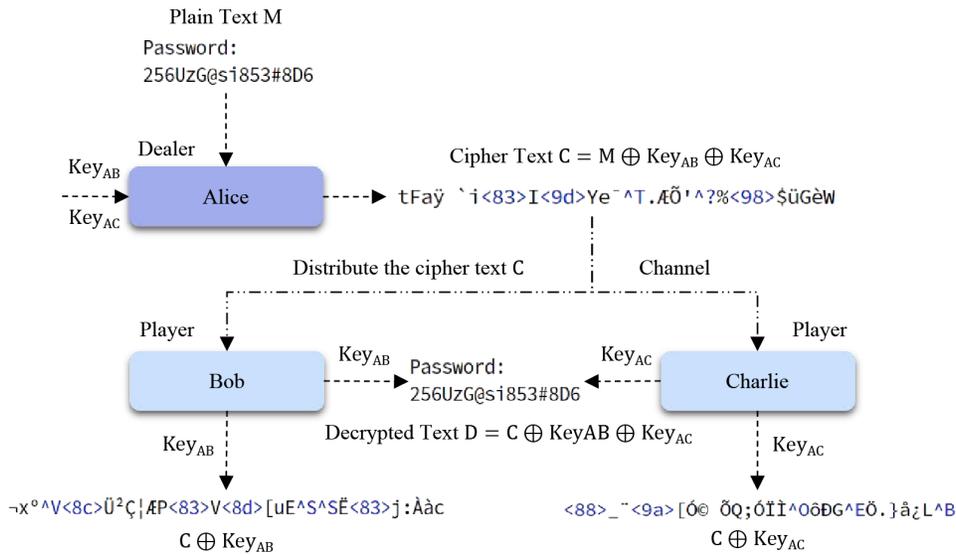

**Fig. 6.** Experimental procedure and results of three-user quantum secret sharing. The dealer, Alice, encrypt the secret message and transmits the encrypted message to two players, Bob and Charlie, via the classical channels. Bob and Charlie can collaborate with each other to reconstruct the secret message.

Unlike numerous QSS protocols that depend on the sequential transmission of quantum states between players, with each player encoding their data into the states using quantum transformations [61-65], the proposed strategy is capable of generating a set of unconditionally secure keys between any two users. This enables the direct establishment of a QSS network that empowers a dealer to share secret messages with selected players, consequently enhancing the practical feasibility and commercial potential of QSS. Figure 6 illustrates the implementation of a three-user QSS network. In this network, the dealer, Alice, encrypts a secret message M by performing $M \oplus \text{Key}_{AB} \oplus \text{Key}_{AC}$, and transmit the resulting encrypted message C to two players, Bob and Charlie, via the classical channels. By sharing their secret keys and decrypting the received message using the equation $C \oplus \text{Key}_{AB} \oplus \text{Key}_{AC}$, the players are able to reconstruct the secret message, while any attempt to obtain useful information without cooperation between the players will be futile.

## 6. Materials and Methods

In experiments, the highly versatile modal birefringence in BRW enabled the PM of the three modalities[66, 67]. Here we consider the properties of biphotons generated from a 4 mm non-ideal quarter-wavelength BRW with high modal overlap in the process of degenerate type-II SPDC[68], i.e., one TE-polarized pump photon at frequency $\omega_p$ is converted into a pair of cross-polarized signal and idler photons at frequency $\omega_s$ and $\omega_i$, respectively, with $\omega_p = \omega_s + \omega_i$. The sample contains a core $Al_{xc}Ga_{1-xc}As$ layer with $x_c$=0.17 and thickness of 230 nm, sandwiched in a six-period Bragg stack made of alternative 127 nm high ($Al_{0.28}Ga_{0.72}As$) and 622 nm low ($Al_{0.72}Ga_{0.28}As$) index layers. We grow the sample along the [001] crystal axis with a width of 5.1 μm and an etching deep of 4.15 μm. The waveguide achieves PM by employing bounded TIR modes and quasi-bounded BRW modes, where TIR modes are commonly formed between high- and low-index claddings and BRW modes are guided though transverse Bragg reflections at the interface between core and period claddings (see Supplementary Note 1 for a theoretical analysis and a detailed characterization of the BRW source). The source is pumped by a tunable CW laser, producing degenerate entangled photons centered at 1560.16 nm, which is aligned with the International Telecommunication Union's (ITU) grids. The polarization entanglement fidelity is above 96% within 42 nm bandwidth (see Supplementary Note 2). The temperature of the waveguide can be stabilized by a temperature controller. The wavelength allocation module is realized either with a WSS (model Finisar 16000s) or a coarse WDM unit followed by multiple DWDM filters with 100/200 GHz spacing, depending on the spectral region of interest. Each detection module consists of an InGaAs avalanche SPD, which has 8.5%/10% detection efficiency, 800Hz dark count rates, and a 17 μs dead time.

## 7. Discussion

In summary, we have designed and fabricated a non-ideal quarter-wavelength BRW with high figure-of-merit modal overlap between interacting fields for photon-pair generation. The broadband polarization entanglement with high fidelity in the telecom band can be generated directly due to the little material birefringence. With this source, we showcase the potential of subspace coding for mitigating noise in QNs. We have analyzed the noise resilience theoretically and experimentally, showing that the QBER decreases significantly as the number of frequency bins N increases, particularly in highly noisy channels. This implies that quantum correlation can effectively overcome physical noise and enhance the channel capacity, even for extreme noise conditions that would otherwise preclude quantum communication. Furthermore, we have demonstrated a fully connected three-user noisy network by multiplexing 3 pairings of DWDM channels such that polarization entanglement can be shared between every possible two-party link. The broad spectrum combined with considerable brightness, make the BRW entangled source highly promising for large-scale, wavelength multiplexed, fully connected QNs. Additionally, by employing time-multiplexing in the polarization analysis module, we can reduce the number of SPDs needed to one per user in the network. Finally, we have performed image encryption and QSS with the secure keys generated among each user by considering the finite key effect. With the advancements of fabrication technology, it is possible to integrate laser and wavelength demultiplexing/multiplexing module onto a single chip, thereby reducing the complexity and connection losses associated with off-chip individual optical components. This work lays the foundation for a turn-key solution for large-scale QNs that are compatible with fiber optical communication.

**Acknowledgements.** We thank Dr. F. Appas for instructive suggestions, and Dr. X.M. Gu for feedback on this manuscript.

**Funding.** This research is supported by the National Natural Science Foundation of China (Grant No. 12274233, 12174187, 62288101). Cheng Qian acknowledges financial support from the Postgraduate Research & Practice Innovation Program of Jiangsu Province (SJCX23_0569).

# Supplementary Materials

## 1. Theoretical analysis of the Bragg reflection waveguide (BRW) quantum source

The quarter-wave BRW structure has been extensively studied in theoretical work due to its ability to confine guided modes in the core, for example [1-3]. Here we will handle a non-ideal quarter-wavelength $Al_xGa_{1-x}As$/GaAs BRW with high figure-of-merit modal overlap for type-II degenerated spontaneous parametric down-conversion (SPDC) processes, i.e., $TE_{BRW} \rightarrow TE_{TIR}+TM_{TIR}$. We first consider a one-dimensional (1D) BRW structure as shown in Fig. S1a, where the core layer has a small composition $x_c$ and thickness $t_c$, which is different from the usual design. The cladding layers consist of compositions $x_a$ and $x_b$ ($x_c<x_a<x_b$) with thickness $a$ and $b$ respectively. The core layer has refractive index $n_c$ and thickness $t_c$, while the cladding layers consist of index $n_a$ and $n_b$ ($n_c >n_a > n_b$). The period of Bragg reflector is $\Lambda = (a + b)$ and the waveguide is symmetric about the center of the Bragg stack position, i.e., x=0. We take the lowest-order TE and TM modes on the x≥0 region as examples, and assume their propagation along the z direction, then the electric field can be expressed as [4]

$$E_y^{TE}(x) = \begin{cases} C_1^{TE} \cos(k_c x), & 0 \leq x \leq \frac{t_c}{2}, \\ C_2^{TE} E_K^{TE}(x - \frac{t_c}{2}) \exp[iK^{TE}(x - \frac{t_c}{2})], & x > \frac{t_c}{2}, \end{cases}$$

(S1)

$$H_y^{TM}(x) = \begin{cases} C_1^{TM} \cos(k_c x), & 0 \leq x \leq \frac{t_c}{2}, \frac{n_1^2 k_2}{n_2^2 k_1} < 1, \\ C_1^{TM} \sin(k_c x), & 0 \leq x \leq \frac{t_c}{2}, \frac{n_1^2 k_2}{n_2^2 k_1} > 1, \\ C_2^{TM} H_K^{TM}(x - \frac{t_c}{2}) \exp[iK^{TM}(x - \frac{t_c}{2})], & x > \frac{t_c}{2}, \end{cases}$$

(S2)

where $K^i$ (i=TE, TM) is the Block wave vector, $k_c = k_0\sqrt{n_c^2 - n_{eff}^2}$ is the transverse wave vector in the core layer. $k_0$ and $n_{eff}$ denote the free-space wave vector and modes effective refractive index. The electric field of TM mode can be solved by $E_x^{TM}(x) = \beta/\omega n^2 H_y^{TM}(x)$. According to the Floquet theorem, the fields $E_K^{TE}(x)$ and $H_K^{TM}(x)$ are periodic with Bragg reflector period $\Lambda$, therefore the electric and magnetic fields in the $n$th unit cell are

$$E_n^{TE}(x) = \begin{cases} a_{En}^{TE} \cos[k_a(x - \frac{t_c}{2} - n\Lambda)] + b_{En}^{TE} \sin[k_a(x - \frac{t_c}{2} - n\Lambda)], \\ \qquad n\Lambda \leq x - \frac{t_c}{2} \leq n\Lambda + a, \\ c_{En}^{TE} \cos[k_b(x - \frac{t_c}{2} - n\Lambda - a)] + d_{En}^{TE} \sin[k_b(x - \frac{t_c}{2} - n\Lambda - a)], \\ \qquad n\Lambda + a \leq x - \frac{t_c}{2} \leq (n+1)\Lambda, \end{cases} \quad (S3)$$

$$H_n^{TM}(x) = \begin{cases} a_{Hn}^{TM} \cos[k_a(x - \frac{t_c}{2} - n\Lambda)] + b_{Hn}^{TM} \sin[k_a(x - \frac{t_c}{2} - n\Lambda)], \\ \qquad n\Lambda \leq x - \frac{t_c}{2} \leq n\Lambda + a, \\ c_{Hn}^{TM} \cos[k_b(x - \frac{t_c}{2} - n\Lambda - a)] + d_{Hn}^{TM} \sin[k_b(x - \frac{t_c}{2} - n\Lambda - a)], \\ \qquad n\Lambda + a \leq x - \frac{t_c}{2} \leq (n+1)\Lambda, \end{cases} \quad (S4)$$

where $a_{jn}^{TE/TM}$ ($b_{jn}^{TE/TM}$) and $c_{jn}^{TE/TM}$ ($d_{jn}^{TE/TM}$) (j= E/H), are the incident (reflected) amplitudes in $n_a$ and $n_b$ regions respectively. With the transfer matrix method and continuous condition at the boundaries, one can obtain

$$\begin{pmatrix} a_{En+1}^{TE} \\ b_{En+1}^{TE} \end{pmatrix} = \begin{bmatrix} \cos(k_b b) & \sin(k_b b) \\ -\frac{k_b}{k_a}\sin(k_b b) & \frac{k_b}{k_a}\cos(k_b b) \end{bmatrix} \begin{pmatrix} c_{En}^{TE} \\ d_{En}^{TE} \end{pmatrix}$$

$$= \begin{bmatrix} \cos(k_b b) & \sin(k_b b) \\ -\frac{k_b}{k_a}\sin(k_b b) & \frac{k_b}{k_a}\cos(k_b b) \end{bmatrix} \begin{bmatrix} \cos(k_a a) & \sin(k_a a) \\ -\frac{k_a}{k_b}\sin(k_a a) & \frac{k_a}{k_b}\cos(k_a a) \end{bmatrix} \begin{pmatrix} a_{En} \\ b_{En} \end{pmatrix}$$

$$= \begin{bmatrix} A_E^{TE} & B_E^{TE} \\ C_E^{TE} & D_E^{TE} \end{bmatrix} \begin{pmatrix} a_{En}^{TE} \\ b_{En}^{TE} \end{pmatrix} \quad (S5)$$

and

$$\begin{pmatrix} c_{Hn}^{TM} \\ d_{Hn}^{TM} \end{pmatrix} = \begin{bmatrix} \cos(k_a a) & \sin(k_a a) \\ -\frac{n_b^2 k_a}{n_a^2 k_b}\sin(k_a a) & \frac{n_b^2 k_a}{n_a^2 k_b}\cos(k_a a) \end{bmatrix} \begin{pmatrix} a_{Hn}^{TM} \\ b_{Hn}^{TM} \end{pmatrix}, \quad (S6)$$

$$\begin{pmatrix} a_{Hn+1}^{TM} \\ b_{Hn+1}^{TM} \end{pmatrix} = \begin{bmatrix} \cos(k_b b) & \sin(k_b b) \\ -\frac{n_a^2 k_b}{n_b^2 k_a}\sin(k_b b) & \frac{n_a^2 k_b}{n_b^2 k_a}\cos(k_b b) \end{bmatrix} \begin{pmatrix} c_{Hn} \\ d_{Hn} \end{pmatrix}$$

$$= \begin{bmatrix} \cos(k_b b) & \sin(k_b b) \\ -\frac{n_a^2 k_b}{n_b^2 k_a}\sin(k_b b) & \frac{n_a^2 k_b}{n_b^2 k_a}\cos(k_b b) \end{bmatrix} \begin{bmatrix} \cos(k_a a) & \sin(k_a a) \\ -\frac{n_b^2 k_a}{n_a^2 k_b}\sin(k_a a) & \frac{n_b^2 k_a}{n_a^2 k_b}\cos(k_a a) \end{bmatrix} \begin{pmatrix} a_{Hn}^{TM} \\ b_{Hn}^{TM} \end{pmatrix}$$

$$= \begin{bmatrix} A_H^{TM} & B_H^{TM} \\ C_H^{TM} & D_H^{TM} \end{bmatrix} \begin{pmatrix} a_{Hn}^{TM} \\ b_{Hn}^{TM} \end{pmatrix}. \quad (S7)$$

Furthermore, the periodicity of the electric field can be expressed as

$$\begin{pmatrix} a_{jn+1} \\ b_{jn+1} \end{pmatrix} = \exp(iK_j\Lambda) \begin{pmatrix} a_{jn} \\ b_{jn} \end{pmatrix}. \quad (S8)$$

From Eqs. (S6), (S7) and (S8), we have

$$exp(iK_j\Lambda) = \frac{A_j+D_j}{2} \pm \sqrt{[\frac{A_j+D_j}{2}]^2 - 1}, \tag{S9}$$

and

$$\begin{pmatrix}a_{jn}\\b_{jn}\end{pmatrix} = \begin{pmatrix}B_j\\exp(iK_j\Lambda) - A_j\end{pmatrix} = exp(-inK_j\Lambda)\begin{pmatrix}a_{j0}\\b_{j0}\end{pmatrix}. \tag{S10}$$

By using the continuity of fields at the core-cladding interface, the coefficients in the first layer of the periodical structures are equal to

$$\begin{pmatrix}a_{E0}\\b_{E0}\end{pmatrix} = \begin{pmatrix}\cos(k_c t_c/2)\\-\frac{k_c}{k_1}\sin(k_c t_c/2)\end{pmatrix} \tag{S11}$$

and

$$\begin{pmatrix}a_{H0}\\b_{H0}\end{pmatrix} = \begin{pmatrix}\cos(k_c t_c/2)\\-\frac{n_1^2 k_c}{n_c^2 k_1}\sin(k_c t_c/2)\end{pmatrix}, \frac{n_1^2 k_2}{n_2^2 k_1} < 1$$

$$\begin{pmatrix}a_{H0}\\b_{H0}\end{pmatrix} = \begin{pmatrix}\sin(k_c t_c/2)\\\frac{n_c^2 k_1}{n_1^2 k_c}\cos(k_c t_c/2)\end{pmatrix}, \frac{n_1^2 k_2}{n_2^2 k_1} > 1 \tag{S12}$$

Substituting Eqs. (S11) and (S12) into (S10), the mode dispersion equations are given by

$$\frac{1}{k_1}\frac{B_E}{exp(iK_E\Lambda) - A_E} = -\frac{1}{k_c}\cot(\frac{k_c t_c}{2}). \tag{S13}$$

and

$$\frac{n_1^2}{k_1}\frac{B_H}{exp(iK_H\Lambda) - A_H} = -\frac{n_c^2}{k_c}\cot(\frac{k_c t_c}{2}) \quad \frac{n_1^2 k_2}{n_2^2 k_1} < 1$$

$$\frac{n_c^2}{k_c}\frac{B_H}{exp(iK_H\Lambda) - A_H} = \frac{n_1^2}{k_1}\cot(\frac{k_c t_c}{2}) \quad \frac{n_1^2 k_2}{n_2^2 k_1} > 1. \tag{S14}$$

The TIR modes can be regarded as special BRW modes confined at the interfaces, where the effective refractive index ($n_{eff}$) is greater than the index of the surrounding medium ($n_2$). Therefore, the derivations presented above are also applicable to the down converted frequencies. To ensure a reasonable operating wavelength and achieve a large modal overlap while avoiding the possibility of photoluminescence caused by photon absorption, we respectively choose $x_c$ = 0.17, $x_a$ = 0.28, $x_b$ = 0.72, $t_c$ =230 nm, a= 127 nm and b=622 nm in fabricating our real sample. The sample is grown along the [001] crystal axis, with a width of 5.1 μm and an etching depth of 4.15 μm. We define the overlap integral of the different modes involved in the SPDC process as

$$\eta = |\frac{1}{M} \int E_p(x,y) E_s(x,y) E_i(x,y) dx dy| \quad (S15)$$

where $M$ is the normalization factor. Fig. S1B displays the normalized electric field distributions along the structure growth direction with 1D analytical (solid) and cut-lines of 2D COMSOL simulations (dashed). The overlap is 15.57%. The figure demonstrates that the two simulation results are in good agreement, indicating that the 1D eigen equation presented here can provide a rapid guideline for the structure design process.

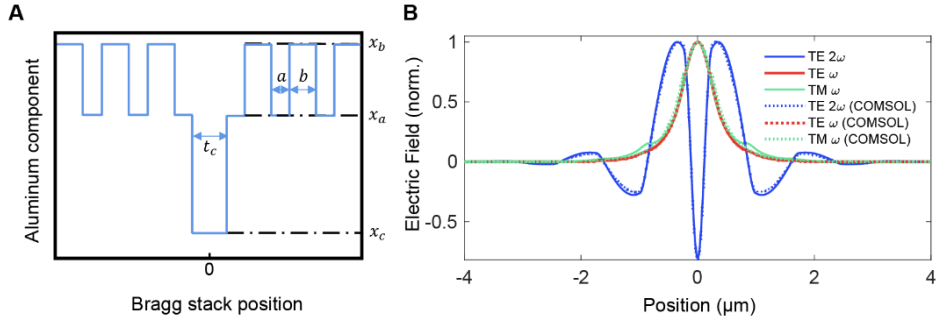

**Fig. S1.** (A) Schematic of the proposed BRW structure. (B) Normalized electric field distribution along the growth direction of the waveguide we designed with 1D solutions (solid) and 2D COMSOL (dashed).

For the degenerated type-II SPDC process in the BRW, a TE-polarized continuous wave laser at frequency $\omega_p$ pumps the waveguide and generates a pair of cross-polarized signal and idler photons at frequency $\omega_s$ and $\omega_i$, respectively, with energy conservation $\omega_p = \omega_s + \omega_i$. The biphoton state $|\Psi\rangle$ at the output of the BRW can be expressed as

$$|\Psi\rangle = |vac\rangle - \frac{i}{\hbar} \int H_I(t) dt \quad (S16)$$

where $H_I$ is the interaction Hamiltonian and can take the form (we define modes TE=H and TM=V in waveguide)

$$H_I \propto \int [E_p^{H(+)} E_s^{V(-)} E_i^{H(-)} + E_p^{H(+)} E_s^{H(-)} E_i^{V(-)}] d\vec{r} + c.c \quad (S17)$$

In the low-gain regime, the signal and idler fields can be quantized as

$$E_j^{m(-)} \propto \int d\,\omega_j\, a_m^+(\omega_j)\, exp(i\beta_j^m z - i\omega_j t), \tag{S18}$$

where j=s, i and m=H, V. Inserting Eqs. (S17) and (S18) into (S16) gives [5]

$$|\Psi\rangle \propto [\int d\,\omega_{sH}\, d\,\omega_{iV}\, f(\omega_{sH}, \omega_{iV}) a^+(\omega_{sH}) a^+(\omega_{iV})$$

$$+ \int d\,\omega_{sV}\, d\,\omega_{iH}\, f(\omega_{sV}, \omega_{iH}) a^+(\omega_{sV}) a^+(\omega_{iH})]|0\rangle$$

$$\tag{S19}$$

where

$$f(\omega_{sH}, \omega_{iV}) = \exp\left[i\frac{(k_{sH}+k_{iV})L}{2}\right] \delta(\omega_{sH} + \omega_{iV} - \omega_{pH})\, sinc\left[\frac{\Delta k_{HV}L}{2}\right],$$

$$\tag{S20}$$

and

$$f(\omega_{sV}, \omega_{iH}) = \exp\left[i\frac{(k_{sV}+k_{iH})L}{2}\right] \delta(\omega_{sV} + \omega_{iH} - \omega_{pH})\, sinc\left[\frac{\Delta k_{VH}L}{2}\right],$$

$$\tag{S21}$$

Where $k_{xy}$ is the waveguide propagation constant for the $y$ polarization of $x$ photon. $\Delta k_{HV}$ and $\Delta k_{VH}$ are the wave-vectors mismatch between the pump and converted photons, L is the length of the waveguide. By using a dichroic beam splitter, signal and idler photons can be separated into different paths, e.g. path 1 and path2. Then the state is given by

$$|\Psi\rangle \propto \int d\,\omega_{sH}\, d\,\omega_{iV}\, f(\omega_{sH}, \omega_{iV})|\omega_s, H\rangle_1 |\omega_i, V\rangle_2$$

$$+ \int d\,\omega_{sV}\, d\,\omega_{iH}\, f(\omega_{sV}, \omega_{iH})|\omega_s, V\rangle_1 |\omega_i, H\rangle_2$$

$$\tag{S22}$$

The state is maximally entangled if $f(\omega_{sH}, \omega_{iV})$ and $f(\omega_{sV}, \omega_{iH})$ are identical, and the phase difference between them is nearly constant. After tracing over frequency terms, the polarization state density matrix is reduced to

$$\rho = 1/N[|\alpha_1|^2|HV\rangle\langle VH| + \alpha_1\alpha_2^*|HV\rangle\langle VH| + \alpha_2\alpha_1^*|VH\rangle\langle HV| + |\alpha_2|^2|VH\rangle\langle HV|], \tag{S23}$$

where

$$\alpha_1 = \int d\,\omega_{sH}\, d\,\omega_{iV}\, f(\omega_{sH}, \omega_{iV}), \tag{S24}$$

$$\alpha_2 = \int d\,\omega_{sV}\, d\,\omega_{iH}\, f(\omega_{sV}, \omega_{iH}), \tag{S25}$$

and $N = |\alpha_1|^2 + |\alpha_2|^2$ is the normalization constant. Owing to the small birefringence, $f(\omega_{sH}, \omega_{iV})$ can overlap with $f(\omega_{sV}, \omega_{iH})$ over a wide spectral range. Using a bandwidth filter to tailor the broadband SPDC photons is a flexible way to directly generate the polarization-entangled state without any additional walk-off compensation.

## 2. Details of the device and experiment

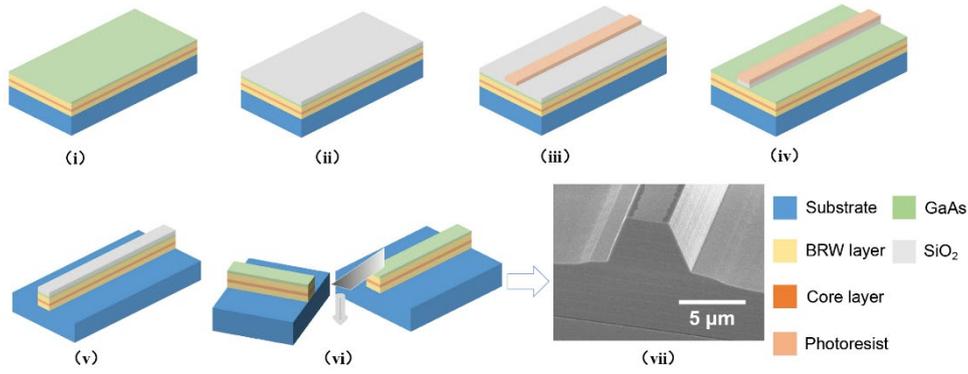

**Fig. S2.** Six-step fabrication process of the BRW source. (i) The device is fabricated on a semi-insulating GaAs substrate. BRW layers are grown on the chip by molecular beam epitaxy. (ii) A thickness of 1 μm $SiO_2$ cladding is deposited via plasma enhanced chemical vapor deposition. (iii) The waveguide pattern is defined by lithography along GaAs crystal orientation [110]. (iv) $SiO_2$ is etched away in a reactive ion etching step with fluorine gas with patterned photoresist as etching mask. (v) Then AlGaAs waveguide ridge was etched by carefully tuning chlorine gas to ensure the planeness of the sidewall. (vi) Finally, the sample is cut along the cleavage plane with smooth edges. (vii) The scanning electron microscope image of the fabricated BRW sample.

The six fabrication steps of the BRW source are depicted in Fig. S2. To evaluate the performance of the source, we employ an experimental setup shown in Fig. S3A. A tunable laser centered at 780.08 nm is polarized and coupled into the chip using lensed fibers. Entangled photons emerging from the chip are filtered by off-chip 980/1550 dense wavelength division multiplexers (DWDMs) to remove the residual pump and are finally selected by DWDMs and detected by two free running single photon detectors (SPDs) with 10% and 8.5% detection efficiency, 800 Hz dark count rates, and a dead time of 17 μs respectively. The detector signals are collected by a field programmable gate array-based

timetag device. The single photon counts in signal and idler channels can be expressed as

$$N_s^m = S_s^t + DC_s + N_{aps} + \zeta P^\alpha \eta_s + N_{rams}\eta_{dets} \tag{S26}$$

$$N_i^m = S_i^t + DC_i + N_{api} + \zeta P^\alpha \eta_i + N_{rami}\eta_{deti} \tag{S27}$$

where

$$S_s^t = B\eta_s \qquad S_i^t = B\eta_i \tag{S28}$$

Here, B is the source brightness in the measured bandwidth, $\eta_s$ ($\eta_i$) represents the heralding efficiency of signal (idler) photon. $DC_s$ ($DC_i$) and $N_{aps}$ ($N_{api}$) are the dark count rate and afterpulse of the detector. The probability of detecting an afterpulse $\eta_{AP}$ is a function of the total detection counts. $\zeta$ is the photoluminescence generation efficiency[6]. $N_{rams}$ and $N_{rami}$ are Raman noise photons generated before the detectors if there is a classical light coexistence with quantum light in the fiber.

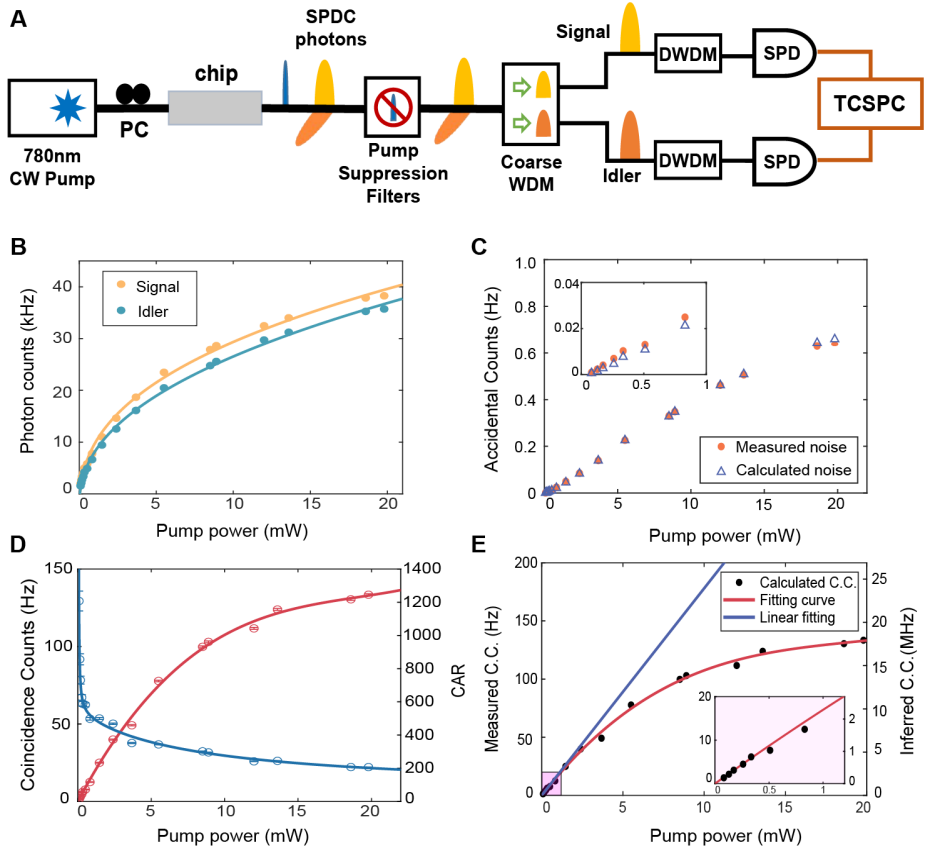

**Fig. S3.** Characterization of the source. (A) Schematic of the experimental setup. (B) Signal and idler photon count rates versus input pump power. The difference between the two count rates is mainly due to their different detector efficiencies. The detectors tend to saturate at high pump power. (C) Measured (circles) and calculated (triangles)

accidental coincidence counts (ACC) versus pump power. The measured ACC are estimated by three time windows away from the coincidence peak. (D) Measured CC and coincidence to accidental coincidence (CAR) versus pump power. The error bars are calculated by assuming a Poissonian distribution. (E) The photon pair generation rate inferred by taking into account the losses and bandwidth of each photon.

After considering the detector dead time ($\tau_{dead}$), the measured single photon rate can be written as

$$S_s^m = N_s^m \cdot \frac{1}{1+N_s^m * \tau_{dead}} \qquad S_i^m = N_i^m \cdot \frac{1}{1+N_i^m * \tau_{dead}} \tag{S29}$$

Therefore, we can write the measured coincidence counts ($CC^m$) and accidental coincidence counts ($CC^{acc}$) as [7]

$$CC^m = B\eta_s\eta_i + CC^{acc} \tag{S30}$$

$$CC^{acc} = S_s^m S_i^m t_{CC} \tag{S31}$$

where $t_{CC} = 0.5 ns$ is the temporal width of the coincidence window. Fig. S3B illustrates the measured and simulated count rates of signal and idler photons under different pump powers. Ideally, $S_s^m$ ($S_i^m$) depends linearly on the pump power. However, as the pump power increases, the detector tends to saturate, causing the rate to deviate from the linear curve. We fit the measured data according to Eqs.(S26-S31) with the heralding efficiencies $\eta_s = 0.01260$, and $\eta_i = 0.01134$, which can be decomposed as follows: the chip-fibre loss is 5.5 dB (measured by launching light at 1550 nm and collecting the output power from the source), the insertion losses over the full band of the off-chip WDMs are 3-3.5 dB, and the detection losses are 10~10.7 dB. The parameter values obtained from the fits are $\zeta \simeq 0.46$ MHz/mW/nm, α=0.6, B=0.14 MHz/mW/nm and $\eta_{AP} = 3\% \sim 7\%$, depending on the detected total counts. The measured (obtained by averaging three time windows counts away from the coincidence peak) and calculated (according to Eq. (S31)) $CC^{acc}$ is shown in Fig. S3C. Fig. S3D shows the measured coincidence results and the coincidence-to-accidental ratio (CAR) under different pump power. By considering the losses and bandwidth (21 nm with above 96% fidelity, see Fig. S4B) of each photon, we calculated the generation rate of photon-pairs shown in Fig. S3E. The inferred polarization entangled generation rate is approximately 2.4 MHz/mW, which is consistent with the above fitting value. By taking into account the 80% reflectivity of the facet for BRW mode [8], the rate can reach 12 MHz/mW, which is an ultra-bright on-chip polarization entangled source. The inset shows the rate changes linearly with the low pump power before detectors saturation.

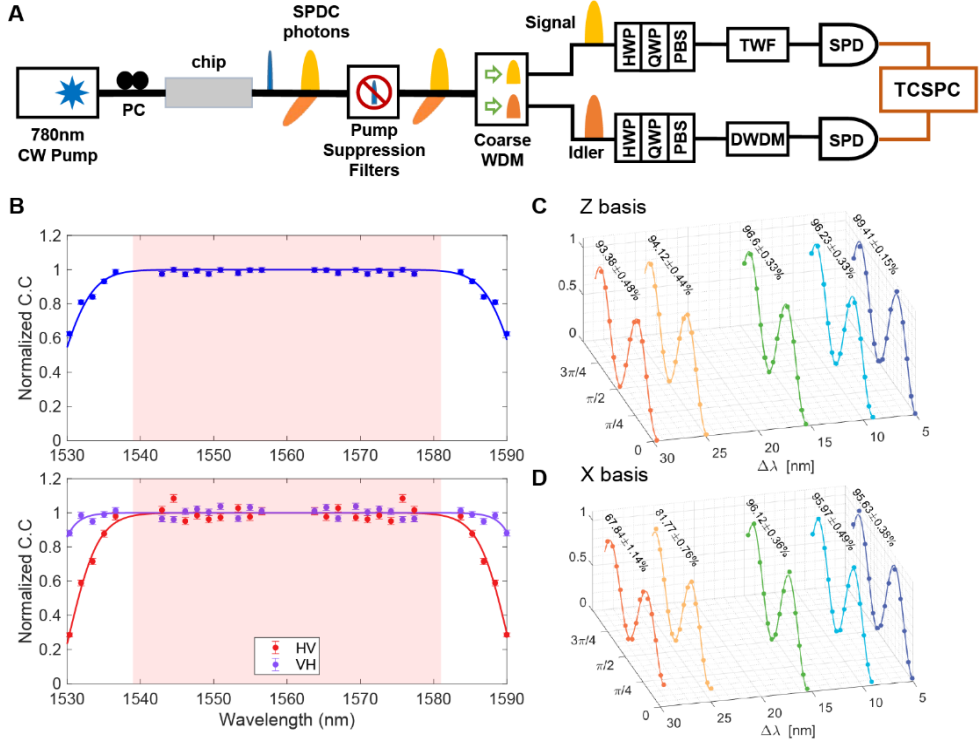

**Fig. S4.** Characterization of the broadband polarization entanglement. (A) Schematic of the experimental setup. (B) Normalized count rates over 60 nm bandwidth with (up) and without (down) projecting the photons to H/V states. Interference curves for the Z basis (C) and X basis (D) as a function of the detuning to degeneracy wavelength. The error bars are calculated by a Poissonian distribution. The dots are the measured data, and the curves are drawn to guide the eye.

To demonstrate the broadband SPDC in the waveguide, we measure the continuous spectra for the signal and idler photons. In the experimental setup shown in Fig. S4A, a CW laser is coupled into the chip. The photons coupling out from the chip are filtered by 200 GHz dense WDM (DWDM) filters and a tunable wavelength filter (TWF). Fig. S4B show the normalized CC rates with (up) and without (down) projecting the photons to H/V states. It can be seen that the generated "HV" and "VH" biphoton spectrums are highly overlapped over 42 nm around the degeneracy, indicating a high degree of entanglement within this spectrum. In order to quantify the correlations of the photon pairs, Fig. S4C and D display interference fringes for the H/V (Z) and A/D (X) bases. The lower bound on entanglement fidelity, which is calculated by the averaged the visibilities in Z and X bases [9], can exceed 0.96 (0.8) over a 42nm (60nm) range. Several monolithically integrated polarization entanglement sources are summarized in Table S1, where the bandwidth and brightness of [10] are deduced from the experimental results. It can be seen that the BRW waveguide can be considered as a leading on-chip source for generating polarization entanglement with both high rate and broad bandwidth.

**Table S1. Comparison of fully polarization entanglement sources.**

| Reference | Material | Bandwidth (nm) | Brightness (MHz/mW or MHz/mW$^2$) | Bell-state fidelity (in %) |
|---|---|---|---|---|
| [10] | Silicon | <2 | <0.558 | 91 |
| [5] | Periodically poled silica fiber | 90 | 1.4 | 98.7 |
| [11] | Ti:LiNbO3 | 2.3 | 28 | 94.5 |
| [8, 12] | AlGaAs | 64 | 3.4 | >85 |
| This work | AlGaAs | 40 | 12 | >96 |
|  |  | 60 | 18 | >80 |

## 3. Realistic channel noise simulation

In this section, we model the advantage of frequency bins number $N$ over a noisy channel, which is a realistic scenario for fiber communication networks. The noise is mainly caused by crosstalk from the classical light when the used DWDM has insufficient isolation, and Raman scattering [13]. To mitigate non-adjacent channel crosstalk, we use appropriate filters at quantum receivers, as we adopt a classical light wavelength is away from the quantum signals. The number of forward Raman scattering photons at wavelength $\lambda_{quantum}$ when a classical light at $\lambda_{classical}$ with power $P_{in}$ is launched into a fiber can be expressed as

$$N_{ram} = P_{in}e^{-\alpha L}L\Delta\lambda \cdot \lambda_{quantum} \cdot \beta(\lambda_{classical}, \lambda_{quantum})/(2hc) \tag{S39}$$

where $\alpha, L$ and $\Delta\lambda$ are, respectively, the fiber attenuation coefficient, the fiber length, and the bandwidth of the filter. $h$ and $c$ are the Plank constant and light speed.

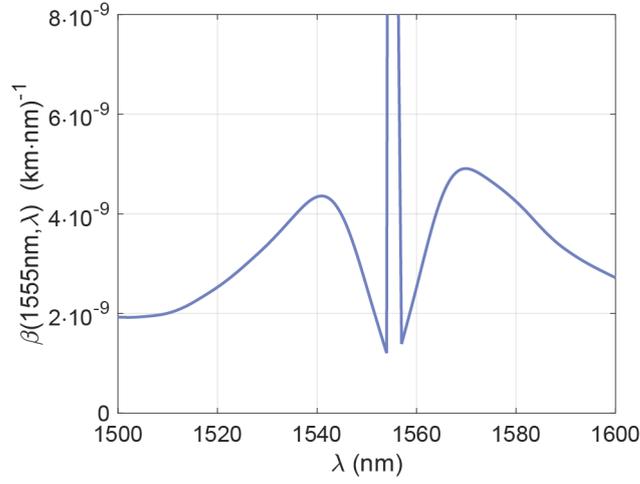

**Fig. S5.** Measured Raman cross section in a single mode fiber with respect to a central wavelength of 1555 nm.

$\beta(\lambda_{classical}, \lambda_{quantum})$ is the Raman cross section, can be written as [14]

$$\beta(\lambda_{classical}, \lambda_{quantum}) = \left(\frac{\lambda_\delta}{\lambda_{quantum}}\right)^4 \cdot \beta(1555\text{nm}, \lambda_\delta) \tag{S40}$$

with

$$\frac{1}{1555\text{nm}} - \frac{1}{\lambda_\delta} = \frac{1}{\lambda_{classical}} - \frac{1}{\lambda_{quantum}} \tag{S41}$$

$\beta(1555\text{nm}, \lambda_\delta)$ is the Raman cross section at $\lambda_\delta$ with respect to a central wavelength of 1555nm in a standard single mode fiber. $\beta(1555\text{nm}, \lambda_\delta)$ is measured and shown in Fig. S5. As we consider the frequency is N-dimensional correlation, the single photon count rate can be rewrite as

$$\tag{S42}$$

$$N_s^m(N) = \frac{S_s^t}{N} + DC_s + N_{aps} + \frac{\zeta P^\alpha \eta_s}{N} + \frac{N_{rams}\eta_{dets}}{N}$$

$$N_i^m(N) = \frac{S_i^t}{N} + DC_i + N_{api} + \frac{\zeta P^\alpha \eta_i}{N} + \frac{N_{rami}\eta_{deti}}{N}$$

The CC and ACC rates for our setup are

$$CC^t = \sum_{i=j=1}^{N} \frac{B\eta_s\eta_i}{N} \tag{S43}$$

$$CC^{acc} = \sum_{i=j=1}^{N} N_s^m(N) \cdot \frac{1}{1 + N_s^m(N) * \tau_{dead}} \cdot N_i^m(N) \cdot \frac{1}{1 + N_i^m(N) * \tau_{dead}} \cdot t_{CC}$$

$$+ \sum_{\substack{i,j=1 \\ i \neq j}}^{N} N_s^m(N) \cdot \frac{1}{1+N_s^m(N)*\tau_{dead}} \cdot N_i^m(N) \cdot \frac{1}{1+N_i^m(N)*\tau_{dead}} \cdot t_{CC} \tag{S44}$$

And the error rate can be given by

$$E = \frac{CC^{err}}{CC^m} = \frac{CC^t e^{pol} + \frac{1}{2}CC^{acc}}{CC^t + CC^{acc}} \tag{S45}$$

where $e^{pol}$ is the polarization error probability, which can be deduced from the interference visibility V as (1-V)/2 for low gain. As shown in Fig. S6, we calculate the signal-to-noise ratio (SNR) for different classical light power as a function of frequency bins according to our experimental setup. These figures clearly demonstrate the performance can be significantly increased by improving N.

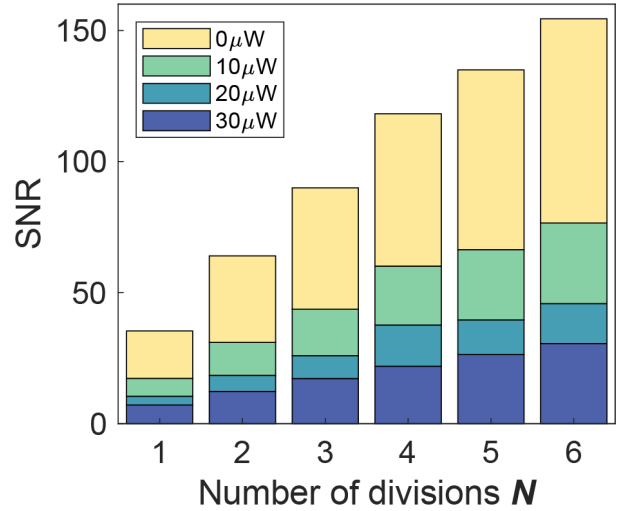

**Fig. S6.** Theoretical results of the signal-to-noise ratio for different classical light power as a function of frequency bin number.

## 4. Polarization analysis and calibration

The polarization analyzer hold by each user is composed of two nested interferometers as shown in Fig. S7. For ease of illustration, we depict light travelling in free space interferometers. The photons are randomly routed and projected to H/V or A/D bases. The basis-dependent delay is introduced by the outer interferometer, and the polarization-dependent shift is introduced by the unbalanced polarization maintaining interferometer (UPMI). This enables the transfer of polarization measurement to the time when photons arrive at the detector. Although the polarization drift can occur due to the random birefringence effects in fibers, active polarization stabilization is generally not required in experiments since all the fibers are located indoors or underground [15, 16]. However, manual compensation is needed before each experiment. Following a procedure similar to [17], we first sweep the telecom-band laser wavelength to enable the phase-matching of degenerated type-I process in the waveguide for second harmonic generation, i.e., $TE_\omega + TE_\omega \rightarrow TM_{2\omega}$. Based on this, the polarization of the telecom-band light emitted from the chip is $TE$. This is equivalent to insert a polarizer after the source that only permits H-polarized photons to pass through. We then adjust the fiber polarization controller (FPC) to minimize the V-polarized counts after the polarization analyzer. However, this only allows compensation of H and V polarization individually, and there is an unknow phase between them. This can be compensated by turning the quarter-wave plate (QWP) and half-wave plate (HWP) to maximize the counts. After that, the effective D/A base is defined.

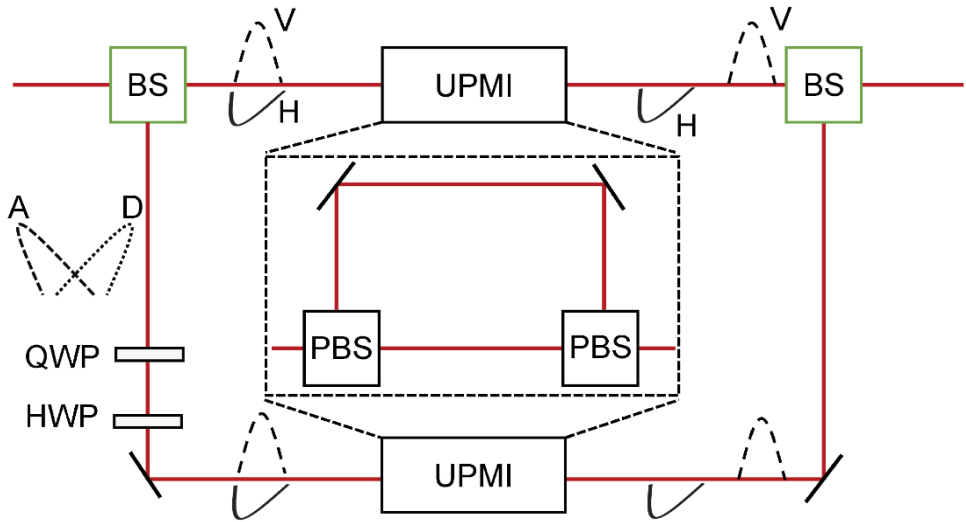

**Fig. S7.** Polarization to time-bin transfer setup. BS, beam splitter; UPMI, unbalanced polarization-maintaining interferometer; PBS, polarizing beam splitter; HWP, half-wave plate; QWP, quarter-wave plate.

## 5. Error correction

After key-sifting, three users reconcile their keys and distill the final secure keys through privacy amplification. During the reconciliation protocol, we adopt low-density parity-check (LDPC) codes for error correction, which is attractive because of its capacity-approaching performance and low-complexity iterative decoding combined with the belief-propagation (BP) algorithm [18, 19]. BP can provide optimum decoding when the LDPC code is cycle-free, so it is important to reduce the impact of cycles. Algorithms based on progressive-edge-growth (PEG) pattern [20-22] is the main kind of LDPC code construction algorithms used to create codes with large girth. In PEG-based algorithms, edges are greedily added into the LDPC code to maximize the length of the current shortest cycle such that the algorithms manage to make the girth larger. For our experiment, the QBER ranges from 4.955% to 10% is divided into 8 parts, and each has a code rate (for example, in the BER range of 4.955% to 5.3655%, errors should be corrected using a code rate of 0.67). In this way, the average correction efficiencies are 1.1004, 1.0986 and 1.0977 respectively. Fig. S8 shows the error correction efficiency of each pair users for 150 data blocks.

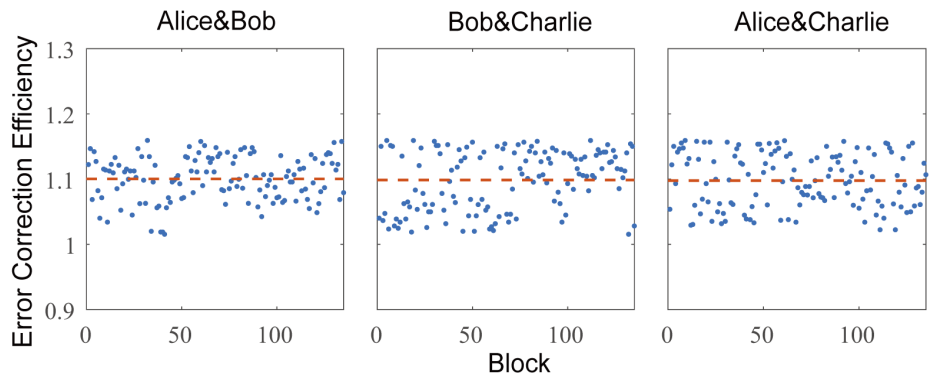

**Fig. S8.** Distribution of LDPC-based error correction efficiencies for the data blocks.

## 6. Image encryption and decryption

Numerous studies have utilized quantum keys to encrypt images through directly performing XOR operations [23, 24]. However, despite the unconditional security provided by quantum keys, this encryption strategy lacks resistance against image-specific attacks due to the high redundancy and correlation among adjacent pixels in the original image. For example, attackers can employ cropping attacks to disrupt the integrity of the encrypted image, thereby preventing legitimate users from retrieving the correct information [25, 26]. As illustrated in Fig. S9, the original image (Fig. S9(a)) is encrypted by performing XOR operations between the quantum keys and the original pixels. During transmission of the encrypted image (Fig. S9 (b)) to the receiver through the classical channel, an attacker executes a cropping attack. Although the receiver can decrypt the cropped image (Fig. S9 (c)) and obtain the decrypted image (Fig. S9 (d)), the core information of the original image becomes obscured.

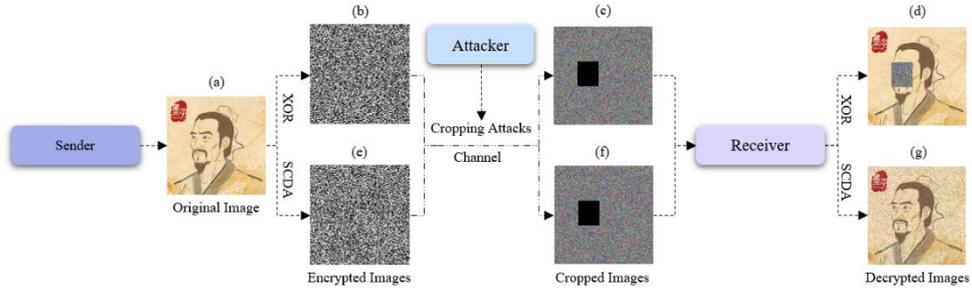

**Fig. S9.** Resistance to cropping attacks. (a) a picture of Micius, (b) encrypted image obtained by directly performing XOR operations between quantum bits and original pixels, (c) cropped image of (b), (d) decrypted image of (c), (e) encrypted image obtained by performing three rounds of confusion and two rounds of diffusion operations on the original image using the quantum bits, (f) cropped image of (e), (g) decrypted image of (f).

An exceptional image encryption algorithm should be capable of recovering the original image with high visual quality, even if the encrypted image is subjected to a cropping attack. Thus, the Shannon Confusion-Diffusion Architecture (SCDA) to encrypt the original image. In our experiment, the sender performs three rounds of confusion operations, followed by two rounds of diffusion and confusion operations to encrypt the original image. For confusion, a discrete version of Chirikov normal map, as shown below, to shuffle the pixel positions [27].

$$\begin{cases} \alpha = (a + o) \bmod N, \\ \beta = \left(o + s_c \sin\left(\frac{2\pi\alpha}{N}\right)\right) \bmod N, \end{cases} \quad (S46)$$

where $a$ and $o$ represent the coordinate of the pixel, $\alpha$ and $\beta$ represent the new coordinate of the pixel after the confusion operation, $N \times N$ is the resolution of the image, and $s_c$ is referred to as the confusion seed, which is a positive integer generated using the secret keys. The inverse transformation of the confusion operation is given by:

$$\begin{cases} a = \left(\alpha - \beta + s_c \sin\left(\frac{2\pi\alpha}{N}\right)\right) \bmod N, \\ o = \left(\beta - s_c \sin\left(\frac{2\pi\alpha}{N}\right)\right) \bmod N. \end{cases} \quad (S47)$$

For diffusion, the following equation is employed to sequentially modify the pixel values [28]:

$$F_e[a, o] = b \oplus (F_o[a, o] + b) \oplus F_e[a', o'], \quad (S48)$$

where $F_e$ and $F_o$ denote the encrypted and original images, respectively, $F[a, o]$ represents the value of the pixel at coordinate $[a, o]$, $b$ is a byte generated using the secret keys, and $[a', o']$ refers to the previous pixel of $[a, o]$. To perform diffusion operation, the diffusion seed $s_d$ needs to be set to process $F[0, 0]$. The inverse transformation of

diffusion operation is defined as follows:

$$F_o[a, o] = (b \oplus F_e[a, o] \oplus F_e[a', o']) - b. \tag{S49}$$

The encryption process is outlined in detail below:

---
**Algorithm S1: Image encryption**

---
**Input**: quantum secret key: $K$; original image: $F_o$; image dimension: $N \times N$, rounds of confusion operations: $r_c$; rounds of diffusion and confusion operations: $r_c$.
**Output**: encrypted image: $F_c$.
Fetch 16 bits from $K$ to construct the confusion seed $s_c$;
Temp Image $F_t = F_o$;
**For** i = 0; i < $r_c$; i++ **do** //Perform $r_c$ round of confusion operations on the original image
    **For** $a$ = 0; $a$ < $N$; $a$++ **do**
        **For** $o$ = 0; $o$ < $N$; $o$++ **do**
            $\alpha = (a + o) \mod N$; $\beta = (o + s_c \sin(2\pi\alpha/N)) \mod N$;
            $F_c[\alpha, \beta] = F_t[a, o]$;
    $F_t = F_c$;
**For** i = 0; i < $r_d$; i++ **do** //Perform $r_d$ round of diffusion and confusion operations on the shuffled image
    Fetch 8 bits from $K$ to construct the diffusion seed $s_d$;
    **For** $a$ = 0; $a$ < $N$; $a$++ **do**
        **For** $o$ = 0; $o$ < $N$; $o$++ **do**
            Fetch 8 bits from $K$ to construct a byte $b$;
            **If** $a == 0$ **And** $o == 0$ **then**
                $F_c[a, o] = b \oplus (F_t[a, o] + b) \oplus s_d$;
            **Else**
                $F_c[a, o] = b \oplus (F_t[a, o] + b) \oplus F_c[a', o']$;
    $F_t = F_c$;
    **For** $a$ = 0; $a$ < $N$; $a$++ **do**
        **For** $o$ = 0; $o$ < $N$; $o$++ **do**
            $\alpha = (a + o) \mod N$; $\beta = (o + s_c \sin(2\pi\alpha/N)) \mod N$;
            $F_c[\alpha, \beta] = F_t[a, o]$;
    $F_t = F_c$;

---

To facilitate a comparison with image encryption strategy based on XOR operations, Algo. S1 is utilized to encrypt the original image. The resulting encrypted image is depicted in Figure S9 (e), which undergoes the same cropping attacks during transmission via the classical channel. The cropped image is illustrated in Figure S9 (f). Notably, despite the identical portion of the encrypted image being cropped, the contours of the original image remain preserved, as evidenced by the decrypted image plotted in Fig. S9 (g).